\documentclass[number]{elsarticle}
\usepackage{grffile}
 \usepackage{graphics}

\usepackage{graphicx}
\usepackage{caption}
\usepackage{subcaption}

\usepackage{epsfig}

\usepackage{graphicx,epsfig}
\usepackage{amsmath}
\usepackage{amsfonts}
\usepackage{amssymb}
\usepackage{amsthm}
\usepackage{amsbsy}
\usepackage{esint}

\usepackage{color}
\usepackage{enumitem}
\setlist{  
  listparindent=\parindent,
  parsep=0pt,
}

\newcommand{\lp}{\left(}
\newcommand{\rp}{\right)}
\newcommand{\beq}{\begin{equation}}
\newcommand{\eeq}{\end{equation}}

\newcommand{\s}{\textbf{S}}
\newcommand{\f}{\textbf{f}}

\newcommand{\lap}{\Delta}
\newcommand{\na}{\nabla}
\newcommand{\tr}{{\rm tr}\;}

\renewcommand{\d}{\partial}
\renewcommand{\div}{\nabla\cdot}

\newcommand{\dgrad}{\cdot\nabla}

\newcommand{\Wi}{\textit{Wi}}

\renewcommand{\u}{\textbf{u}}
\newcommand{\tu}{\tilde{\u}}
\newcommand{\tp}{\tilde{p}}

\renewcommand{\ss}{\tilde{\s}}

\newcommand{\id}{\mathbf{I}}

\newcommand{\Id}{\mathbb{I}}

\renewcommand{\S}{\mathbf{S}}

\newcommand{\st}{\boldsymbol{\tau}_p}
\newcommand{\stn}{\boldsymbol{\tau}_s}
\renewcommand{\ss}{\boldsymbol{A}}
\newcommand{\sig}{\boldsymbol{\sigma}}
\newcommand{\sr}{\boldsymbol{\dot{\gamma}}}

\begin{document}

\begin{frontmatter}

\title{Equilibrium circulation and stress distribution in viscoelastic creeping flow}

\author[jab]{Joseph.~A.~Biello}
\ead{biello@math.ucdavis.edu}

\author[jab]{Becca Thomases\corref{co}}
\cortext[co]{Corresponding author}
\ead{thomases@math.ucdavis.edu}

\address[jab]{Department of Mathematics, University of California, Davis, CA 95616}

\begin{abstract}
An analytic, asymptotic approximation of the nonlinear steady-state equations for viscoelastic creeping flow, modeled by the Oldroyd-B equations with polymer stress diffusion, is derived.  
Near the extensional stagnation point the flow stretches and aligns polymers along the outgoing streamlines of the stagnation point  resulting in a stress-island, or birefringent strand.  The polymer stress diffusion coefficient is used, both as an asymptotic parameter and a regularization parameter. The structure of the singular part of the polymer stress tensor is a Gaussian 
 aligned with the incoming streamline of the stagnation point;
a smoothed $\delta$-distribution whose width is proportional to the square-root of the diffusion coefficient.  The amplitude of the stress island scales with the Wiessenberg number, and although singular in the limit of vanishing diffusion, it is 
integrable in the cross stream direction due to its vanishing width;  this yields 
 a convergent secondary flow.  The leading order velocity response to this stress island is constructed and shown to be  {\em independent} of the diffusion coefficient in the limit.   The  secondary circulation counteracts the forced flow and has a vorticity jump at the location of the stress islands, essentially expelling the background vorticity from the location of the birefringent strands.
The analytic solutions are shown to be in excellent quantitative agreement with full numerical simulations, and therefore, the analytic solutions elucidate the salient mechanisms of  the flow response to viscoelasticity and the mechanism for instability.
\end{abstract}


\end{frontmatter}


\section{Introduction}
\label{intro}

Viscoelastic flows are found in many important engineering and biological systems. Despite the need to understand 
these flows in a variety of complex situations, analysis of the equations of motion describing viscoelastic fluids, even in the low-Reynolds number regime, is very incomplete. There are many different models depending on the rheology of the fluid, but little is known even for the simplest closed continuum models.  One popular model, the Oldroyd-B model, can be derived from microscopic principles and represents ``Boger" fluids, dilute solutions of polymers immersed in a Newtonian solvent which exhibit normal stress differences but not shear thinning.  This model is used frequently in simulations of viscoelastic fluids even though there is no mathematical well-posedness theory for this system, i.e. it is not known if sufficiently smooth solutions to this system exist for all time, bringing in to question the reliability of any numerical simulation.  

Flows at internal stagnation points (such as the four-roll mill flow or the cross-slot or cross channel flow) pose a particular difficulty for both theoretical investigation and numerical simulations of viscoelastic fluids, as polymers are aligned and stretched, and can create fine features in the flow that are difficult to resolve numerically.  However it is precisely at these points in the flow that interesting dynamics arise.  Instabilities have been found in experiments at internal stagnation points \cite{ATDG2006, soulages2009investigating, liu2012oscillations, haward2013instabilities,sousa2015purely},  and  related numerical instabilities are found in similar geometries \cite{harris1993start, harris1994instabilities, poole2007purely, thomases2009transition, xi2009mechanism, thomases2011stokesian}.  It is unclear what is driving these instabilities, but it is reasonable to conjecture that they are related to the large polymer stresses and stress gradients which accumulate along the incoming and outgoing streamlines of these internal stagnation points. 

The elastic contribution to the total stress can be incorporated into the equations of motion by assuming that the total stress on the fluid, $\sigma=\stn+\st,$ comes from a solvent contribution $\stn$ as well as a polymer contribution $\st.$ In the case of a Newtonian solvent, the total stress is given by  \begin{equation*}\sig=-p\id+\eta_s\sr+\st,\end{equation*} where $\eta_s$ is the Newtonian solvent viscosity, and  $\sr=\left[\nabla \u+\nabla\u^{T}\right]$ is the rate-of-strain tensor.  Assuming conservation of mass and incompressibility the fluid velocity $\u$ satisfies \begin{equation*}\rho\frac{D\u}{Dt}=\div\sigma+\mathbf{f},\;\;\div\u=0,\end{equation*} for density $\rho,$ and body force $\mathbf{f},$ or in the inertialess regime, \beq\label{momcon1}\div \sigma+\f=0,\;\; \div\u = 0.\eeq  In the Oldroyd-B model, the symmetric polymer stress tensor, $\st,$  is advected via the upper-convected derivative and relaxes with a characteristic relaxation 
time $\lambda:$ \begin{equation}\label{ob1}  \st+\lambda\stackrel{\nabla}\st=\eta_p\sr.\end{equation}  Here $\eta_p$ is the polymer viscosity, and the upper-convected derivative is defined by \begin{equation*}\stackrel{\nabla}{\ss}
\equiv\frac{\d \ss}{\d t}+\u\dgrad\ss-\left(\nabla\u\cdot\ss+\ss\cdot(\nabla\u)^{T}\right),\textrm{  where   }(\nabla\u)_{ij}=\frac{\d u^i}{\d x_j}.\end{equation*}

While Boger fluids are used in many experiments of viscoelastic phenomena,  it is not immediately clear that the Oldroyd-B model is a good choice for modeling more general complex fluids. We choose to work with this model due to the generic nature of the upper-convected derivative.  This represents a tensorial material derivative and hence will be found in continuum models which advect a macroscopic elastic stress tensor.  Some other variants to the Oldroyd-B model include the Giesekus \cite{giesekus1966}, Phan-Thien-Tanner
(PTT) \cite{thien1977new}, and FENE-P models \cite{peterlin1961}.
These models arise from different microscale models of the polymers.
All of them introduce a nonlinear relaxation of stress which results
in shear-thinning behavior.  All of the above-mentioned macroscopic models contain the
upper-convected derivative, the dominant source of nonlinearity in the
equations, which leads to many of the difficulties and interesting
phenomena associated with the Oldroyd-B
model \cite{BHAC1980,OP2002,poole2007purely,thomases2009transition,guy2015computational}.
Oldroyd-B is the ``simplest" of these models making it a good model
for our theoretical work.

A simple modification to the Oldroyd-B model, which will yield smooth and bounded stresses \cite{T2011, constantin2012note}, is to add polymer stress diffusion.   The addition of stress diffusion can be derived from the kinetic theory 
 of dumbbells \cite{BHAC1980, Larson1999},  but the stress diffusion coefficient is proportional to the square of the ratio of the bead diameter (or polymer radius of gyration) to the flow length-scale, which even in the context of micro-fluidics is minute (on the size of $10^{-6}$ at most) \cite{KL1989}.  To be useful as a regularization in numerical simulations, artificially large polymer stress diffusion is typically needed \cite{SB1995,T2011}.  However it is useful to note that there is an analytical result \cite{constantin2012note} which proves that {\em any} amount of polymer stress diffusion will maintain a smooth and bounded polymer stress. In this manuscript we use polymer stress diffusion to derive an asymptotic expansion, in orders of the square-root of the stress diffusion coefficient, for solutions to the Oldroyd-B model (at zero Reynolds number) in a simple extensional flow geometry. This solution provides information about the effect of the large stress islands, or birefringent strands, on the resultant flow field.  In particular, we are able to take the limit as the diffusion goes to zero and recover information about the effect of these stress islands on the flow.  Therefore we can determine the first order effect of the stress island on the velocity in the Stokes-Oldroyd-B system.    
 
An important structure of the momentum equation, which we use to guide us,  is that at zero-Reynolds number the velocity is one degree smoother than the stress.  This implies that at extensional points in the flow, where the stress accumulates, the exact value of the stress is not needed to determine the effect on the velocity.  Only the integral of the stress affects the velocity field. The stress island can be approximated by a smoothed Dirac $\delta$-distribution.  Furthermore, when stress diffusion is included in the model, a Gaussian becomes an exact solution of the asymptotic approximation for the stress tensor. 

The Gaussian  has a well-defined integral even in the limit of zero diffusion which  enables us to close the asymptotic expansion  and give a well defined solution for the velocity. The result of the transversely narrow and sharply peaked  stress distribution is a dip in the velocity whose magnitude is independent of the stress diffusion.  Such a dip in the velocity field has been observed experimentally  \cite{lyazid1980velocity,gardner1982photon} and provides a possible mechanism for the instabilities seen in numerical simulations \cite{harris1993start,harris1994instabilities,poole2007purely,thomases2009transition,xi2009mechanism, thomases2011stokesian}.  Simply stated, the instability mechanism  is due to the fact  that at extensional points in the flow the vorticity is low.   In the low vorticity region, the stress can grow and where the stress is large the vorticity is expelled, leaving a larger area for the stress to begin to oscillate and become unstable.  Boundary layer approximations near extensional stagnation points that depend on the polymer extension length and relaxation time were presented in \cite{rabin1986flow,harlen1990high}.

In what follows we will describe the model and assumptions and derive an asymptotic expansion for the stress and velocity to first order in the stress diffusion coefficient. We conclude by showing that the solutions to our model agree extremely well with numerical simulations.  The model captures both the leading order velocity response, as well as the amplitude of the stress in the birefringent strands.

\subsection{Model}\label{setup} To perform the analysis it is simpler to write Eqs. \eqref{momcon1}-\eqref{ob1} in terms of a conformation tensor, $\s,$ defined by \beq\s=\Wi\xi^{-1}\st+\Id.\label{stressconf}\eeq  The addition of a polymer stress diffusion term, $\nu\lap\S$ is added to the stress advection equation. This is necessary to our analysis, and we perform the asymptotic expansion in orders of the stress diffusion coefficient $\nu$.   In non-dimensional form we write the Stokes-Oldroyd-B equations with polymer diffusion as \begin{gather}
 \Delta \u - \nabla p +\xi \Wi^{-1}\nabla\cdot\S+ \f = 0, \textrm{ and } \div \u=0, \label{momcon}\\
\Wi  \stackrel{\nabla}{\S}+(\S-\Id)=\nu\lap\S. \label{OB} \end{gather} The Weissenberg  number, $\Wi=\lambda/\tau_f,$ is the ratio of the elastic relaxation time to the
characteristic flow time-scale, set by $\f$ which we set to unity, and  $\xi=\eta_p/\eta_s$ is the ratio of polymer to solvent viscosity. 

\subsection{Outline of solution strategy}

The objective of this work is to find an analytic, asymptotic approximation of Eqs. \eqref{momcon1}-\eqref{ob1} at steady state.  Our analytical strategy has a few key steps which exploit both the structure of the upper convective 
derivative and the linearity of the stress feedback on the Stokes equations.    Our steps will proceed as follows.  
\begin{enumerate}
\item We rescale the velocity field by $\Wi$, yielding a factor of $\Wi$ which multiplies the pressure and the force, $\f$.  After the rescaling,   $\Wi$ does not appear in the advection/diffusion equation for the stress.  
\item  In the rescaled variables, we choose  a simple  background flow, $\u$,  to drive the
dynamics of the upper convective derivative, without specifying the force,  $\f$,  which creates this flow.   {\em Crucially} the flow we  choose has  the property $\partial_y u = 0$ everywhere.  Physically speaking, this is a flow whose vorticity is zero near the maximum of the stress island.     In constructing the solution, we will see that the stress feedback on the flow also produces a velocity field whose vorticity vanishes at the maximum of the stress island.   Additionally, the feedback flow tends to expel vorticity from the vicinity of the maximum of the stress.  

The flow we choose is only intended to describe the local structure of a generic flow near a stress island. This allows us to solve the stress equation because when the stress diffusion coefficient is small the stress equation is essentially hyperbolic, and therefore local in the velocity field. 

 Mathematically, a flow with this structure causes the equation for the conformation tensor to 
 decouple into a hierarchy of three inhomogeneous, non-constant coefficient linear PDEs. The first component of the conformation tensor, $S_{11}$, is forced by a constant.  The second component,  $S_{12},$ is forced by the solution for $S_{11},$ and the third component, $S_{22}$, is forced by $S_{12}$.   
 
The Oldroyd-B  model is most physically relevant in the limit of vanishing stress diffusivity, $\nu$.   This motivates an anisotropic scaling of the spatial coordinates typical of boundary layer theories.  The resulting linear PDE can then be solved analytically, thereby providing the profile of the conformation tensor.

\item 
From the form of the Stokes' equation, Eq. \eqref{momcon}, the conformation tensor feeds back onto the flow through its divergence, whose components we define as $Q_1$ and $Q_2$ as follows:    \beq
    \begin{split}
Q_1 &\equiv \partial_x S_{11} + \partial_y S_{12} \\    
    Q_2 & \equiv \partial_x S_{12} + \partial_y S_{22}.
    \label{q_definition}
    \end{split}
    \eeq
However, since the diffusion is small, and  the equations for the components of the tensor break up into a hierarchy of inhomogeneous equations,  we show that only $Q_1$, the component of the stress divergence in the direction of the axis of localization,  is needed to compute the lowest order effect of the stress on the flow.   

We use  $Q_1$ to compute the velocity field which arises as a response to the stress.   For this problem, we use the classical boundary layer matching techniques whereby the flow is computed in the outer and inner regions separately, and then the two solutions are matched.    The inner region corresponds to the layer where the stress divergence, $Q_1$,  is concentrated.  

\item  At this point, having prescribed the total velocity (at least locally near the stress island),  the conformation tensor and the velocity field
induced by the conformation tensor are computed.  Since the components of the conformation tensor are sharply localized in stress islands, they are therefore only affected by the velocity field in the vicinity of this localization.   
By requiring that the total velocity - that due to the forcing plus that due to the stress response - be consistent along the axis of localization of the stress tensor,   we are able to  establish a simple linear equation for the saturated amplitude of the conformation tensor.

\end{enumerate}

\section{Asymptotic approximation of stress islands and their induced flow}

We will first perform all of the calculations in the case of a single stress island in a domain which is infinite in $y$ and periodic in $x$.   The essence of the calculation is captured by this example, which can easily  be generalized to doubly periodic domains for purposes of comparison to numerical simulations in the 4-roll mill geometry.  We assume periodicity in $x$ for simplicity and consider an extensional flow centered at the origin, with incoming streamlines aligned along the $y-$axis and outgoing streamlines aligned along the $x-$axis. 
 In order to create a stress island of finite length along the $x$-axis, it is necessary that the flow turn around at some $|x| >0$, and then point away from the $x$-axis; this will certainly hold in any physical flow.  Furthermore, by choosing a functional form which is separable in $(x,y)$  the calculation for obtaining  the dependence of the conformation tensor, $\S$ on $x$, becomes straightforward.

 \subsection{Rescaling the velocity}
In order to simplify the presentation of our calculation it is convenient to 
rescale the velocity field by $\Wi.$   Let $\tu=\Wi\; \u,$ and  $\tp=\Wi \; p$,
 then the steady state system of equations becomes 
\begin{gather}
 \Delta \tu - \nabla \tp +\frac{1}{2}\nabla\cdot\S + \Wi \; \f = 0, \textrm{ with } \div \tu=0,\label{OB_rs1} \\
\tu\dgrad\S -\left(\nabla\tu~\S+\S~\nabla\tu^T \right)+(\S-\Id)=\nu\lap\S. \label{OB_rs2}
\end{gather} 
We have set the viscosity ratio $\xi$ to $1/2,$ for convenience. 
Since  equation \eqref{OB_rs1} is linear we can split the velocity $\tu=\u_f+\u_s$ where $\u_f$  solves
\beq
\Delta\u_f-\nabla p_f+\Wi \; \f=0
\label{uf}
\eeq
 i.e. $\u_f$ is the response of the flow field to the background forcing.
  and $\u_s$  solves 
  \beq
  \Delta\u_s-\nabla p_s+\frac{1}{2}\div\S=0
  \label{us_def}
  \eeq
   so $\u_s$ is the stress response.  Rather than specify $\f,$ we prescribe the total flow, $\tu$,
    independent of $\Wi$.     
     Clearly from \eqref{uf}, the background flow, $\u_f$, is linear in $\Wi$, while the
     induced flow is linear in the amplitude of the conformation tensor.   
     
Equation \eqref{OB_rs2} is linear in $\S$  and, because of the re-scaling of the velocity field, the Weissenberg number does not appear in this equation.   
  Therefore the forced portion of the flow must add to the stress induced portion of the flow to give a total scaled flow  ($\tu$) which is independent of $\Wi$ near the region where $\S$ is localized.  This means that the re-scaled flow near the stress island is universal, independent of Weissenbeg number.   
  We are free to choose a form of the velocity field, $\tu$, in equation \eqref{OB_rs2} which is valid locally near the stress island, determine the stress profile generated by this velocity field and then determine the flow, $\u_s$, induced by this velocity field.   The induced velocity,  $\u_s$,  will be exponentially localized near the stress island and its functional form will coincide with  $\u_f$ there.     According to equation \eqref{uf}, $\u_f$ is linear in Weissenberg number.   Therefore, the requirement that $\tu$ is independent of Weissenberg number near the stress island will yield a linear relation for the amplitude of the conformation tensor in terms of the Weissenberg number.    
   
 We will write the rescaled total velocity as $\tu = u \hat{i} + v \hat{j}$, so that the equation for the upper convected derivative of the stress \eqref{OB_rs2} is written explicitly as
\beq
\left[u \partial_x + v \partial_y \right] \S  - 
\left[
\begin{array}{cc}
2 \lp \partial_x u \, S_{11} + \partial_y u \, S_{12} \rp &  \partial_x v \, S_{11} + \partial_y u \, S_{22} \\
\partial_x v \, S_{11} + \partial_y u \, S_{22} &    2 \lp \partial_x v \, S_{12} + \partial_y v \, S_{22} \rp
\end{array}
\right]
+ \lp \S - {\bf I} \rp = \nu  \Delta \S.
\label{ad1}
\eeq

\subsection{The total velocity field}

An incompressible 2-D flow can be described in terms of a stream 
function $\tu = u \, \hat{i} + v \, \hat{j} = -\partial_y \psi \,  \hat{i} + \partial_x \psi \, \hat{j}$ with vorticity
$ \omega = \partial_x v - \partial_y u = \Delta \psi$.   
Taking the curl of (\ref{us_def}), the vorticity induced by the stress satisfies the Poisson equation
\beq
\Delta \omega_s = \frac{1}{2} \lp \partial_y Q_1 - \partial_x Q_2 \rp
\label{om_s}
\eeq 
where  the induced vorticity, velocity and stream function are related through $ \omega_s = \partial_x v_s - \partial_y u_s = \Delta \psi_s$.

Consider the simple stream function
\beq
\psi = - y \, \sin(x)
\eeq
whose velocity field is
\beq
\left[ \begin{array}{c}  u \\ v \end{array} \right]
= 
\left[ \begin{array}{c}  \sin(x) \\ - y \, \cos(x) \end{array} \right].
\label{u}
\eeq
The salient  properties of the flow (\ref{u}) are that it has an extensional point at $(x,y) = (0,0)$, its vorticity is proportional to the stream function
\beq
\omega = y \sin(x),
\eeq
(which vanishes along the $x$-axis) and its deformation tensor is
\beq
\left[ \begin{array}{cc}  \partial_x u & \partial_y u \\ \partial_x v & \partial_y v 
 \end{array} \right] 
 =
\left[ \begin{array}{cc}  \cos(x) & 0 \\ y \, \sin(x) & - \cos(x) 
 \end{array} \right] .
 \eeq
 
 The flow (\ref{u}) is not an exact solution of the stationary Stoke-Oldroyd-B equations, but it is a useful canonical flow if one considers how a stress island is generated.   A circulation  like (\ref{u}) moves fluid toward the origin along the $y$-axis and away from the origin along the $x$-axis.   As a consequence of this flow, a  stress island is formed along the $x$-axis.  Since the stress diffusion, $\nu$, is small, the stress island is confined to a thin layer around the $x$-axis.   Through the Stokes equations, the stress island generates a secondary circulation which, in the vicinity of the stress island, tends to counteract the original flow.   However, since the stress island is strongly confined to the $x$-axis, it only responds to the primary ($\u_f$) and secondary ($\u_s$) circulation in its vicinity.    The flow in (\ref{u}) is simply the first term in the Taylor series in $y$ (near $y=0$) and a Fourier series in $x$  of any general  flow, which is also periodic in $x$.   Therefore, locally about the $x$-axis, any periodic flow would have a leading order term proportional to (\ref{u}).   
 
The choice of a  periodic function in $x$ is also not arbitrary.   If the flow was instead $(u,v) \propto (x,-y)$, 
it would not recirculate, and the resulting stress island would be infinitely long and invariant along the $x$-axis.   
%
%

\subsection{Solution of the conformation tensor equations}
Substituting the velocity field in \eqref{u}  to the equations in 
\eqref{ad1}, the equations for the conformation tensor become
\beq
\begin{split}
\sin(x) \, \partial_x S_{11} - y \cos(x) \, \partial_y S_{11}
+ \left[ 1 - 2 \cos(x) \right] S_{11} - 1 & = \nu \Delta S_{11} \\
\sin(x) \, \partial_x S_{12} - y \cos(x) \, \partial_y S_{12}
+ S_{12} -  y \sin(x) \, S_{11} & = \nu \Delta S_{12} \\
\sin(x) \, \partial_x S_{22} - y \cos(x) \, \partial_y S_{22}
+ \left[ 1 + 2 \cos(x) \right] S_{22} - 2 y \sin(x) \, S_{12} - 1 & = \nu \Delta S_{22} 
\label{stress}
\end{split}
\eeq
These equations have a simple structure, three aspects of which are very illuminating.  
They each have an inhomogeneity, the $S_{11}$ equation has a constant, $1$, the
$S_{12}$ equation has $y \sin(x)S_{11}$ and the $S_{22}$ equation has $2y\sin(x) S_{12}-1$.   The first order $y$-derivatives on the left hand
sides are multiplied by $y$,  with  no other functional dependence on $y$.  Therefore, there is no additional $y$ scale associated with transport and
stretching: however, there is a $y$ scale associated with stress diffusion.  
The antisymmetry of the inhomogeneities coupled with the symmetry of the linear operator results in $S_{11}$ and $S_{22}$ being symmetric and  $S_{12}$ being antisymmetric  about the $y$-axis.

\subsubsection{Anisotropic scaling of coordinates}
In the limit that $\nu \rightarrow 0$ we can approximately solve
the conformation tensor equations \eqref{stress}  by stretching the $y$-coordinate
\beq
\frac{y}{\sqrt{\nu}} = Y \Longrightarrow \frac{\partial}{\partial y} = \frac{1}{\sqrt{\nu}}
\frac{\partial}{\partial Y}
\eeq
where we assume that derivatives with respect to $Y$ are $O(1)$.
As in boundary layer theories the $x$-coordinate is not rescaled.  This is tantamount to assuming that $y$-derivatives are $O(\nu^{-\frac{1}{2}})$ larger than $x$-derivatives.   

Performing the rescaling on equations (\ref{stress}) and retaining the lowest order terms in 
$\nu$ removes second derivative terms in $x$ and we find the equation for $S_{11}$ is a linear, inhomogeneous PDE  independent of $S_{12}$ and $S_{22},$
\beq
\sin(x) \, \partial_x S_{11} - Y \cos(x) \, \partial_Y S_{11}
+ \left[ 1 - 2 \cos(x) \right] S_{11} -  \partial^2_{YY} S_{11} = 1.
\label{s11}
\eeq
The equation for $S_{12}$ is also linear, with an inhomogeneity which depends on $ \sqrt{\nu} S_{11},$
\beq
\sin(x) \, \partial_x S_{12} - Y \cos(x) \, \partial_Y S_{12}
+ S_{12}  - \partial^2_{YY} S_{12}  =  Y \sin(x) \, \sqrt{\nu} S_{11}.
\label{s12}
\eeq
Finally, the equation for $S_{22}$ is also linear, with an inhomogeneity which depends on
$\sqrt{\nu} S_{12},$
\beq
\begin{split}
\sin(x) \, \partial_x S_{22} - Y \cos(x) \, \partial_Y S_{22} &
+ \left[ 1 + 2 \cos(x) \right] S_{22}  - \partial^2_{YY} S_{22}  \qquad \\
&   =  2 Y \sin(x) \, \sqrt{\nu} S_{12} + 1.
 \end{split}
 \label{s22}
\eeq
This is typical of boundary layer scaling, and implies that the stress tensor will be elongated along the $x$-axis.  

It is clear that a solution of the system (\ref{s11}) - (\ref{s22}) results in an asymptotic ordering
\beq
O(S_{12}) \sim  \sqrt{\nu} \, O(S_{11})
\quad
{\rm and}
\quad
O(S_{22}) \sim  \nu \, O(S_{11}).
\eeq
However, it is the derivatives of $S_{ij}$, i.e.  $Q_1$ and $Q_2$ from Eq. (\ref{om_s}),  that drive the response of the fluid to the induced stress. 
Since  partial derivatives with respect to $y$ are a factor of $\nu^{-\frac{1}{2}}$ greater
than partial derivatives with respect to $x$,   we  estimate the order of  $Q_1$ (Eq. \ref{q_definition})
by
\beq
\begin{split}
O(Q_1) & \sim O(S_{11}) + \nu^{-\frac{1}{2}}   \, O(S_{12}) \\
& \sim O(S_{11}) + \nu^{-\frac{1}{2}} \nu^{\frac{1}{2}}  \, O(S_{11}) \\
& \sim O(S_{11}),
\end{split} \eeq
which is to say that both terms which define $Q_1$ contribute 
at  the same order of magnitude (in $\nu$)
to the stress induced velocity field.  Similarly for $Q_2$
\beq
\begin{split}
O(Q_2) & \sim O(S_{12}) + \nu^{-\frac{1}{2}}  \, O(S_{22}) \\
 & \sim \nu^{\frac{1}{2}}  O(S_{11}) +  \nu^{-\frac{1}{2}} \nu  \, O(S_{11}) \\
& \sim    \nu^{\frac{1}{2}} O(S_{11}),
\end{split}
\eeq
which is to say that both terms which define $Q_2$ are of the same order, that being $\sqrt{\nu}$ smaller than $Q_1$.   

Comparing the terms on the right hand side of (\ref{om_s}) we find
\beq 
\begin{split}
O \lp \partial_x Q_2 - \partial_y Q_1 \rp
& \sim
O(Q_2) +  \nu^{-\frac{1}{2}} O(Q_1) \\
& \sim
 \nu^{\frac{1}{2}} O(S_{11}) +
 \nu^{-\frac{1}{2}}  O(S_{11})  \\
 & \sim
 \nu^{-\frac{1}{2}}  O(S_{11}),
\end{split}
\eeq
which means that the $Q_2$ contribution to the torque is a factor of $\nu$ smaller than the $Q_1$ contribution in the vorticity equation (\ref{om_s}).  So it suffices to compute 
\beq
\partial_y Q_1 = \partial^2_{xy} S_{11} + \partial^2_{yy} S_{12}
\label{vort}
\eeq
 in order to calculate $\omega_s$, i.e.
 \beq
 \Delta \omega_s = \frac{\partial_y Q_1}{2} + \mathcal{O}(\nu).
 \eeq 
    The  equation for $Q_1$ is
 constructed by taking the $x$-derivative of (\ref{s11}) plus the $y$-derivative of (\ref{s12})
\beq
\sin(x) \, \partial_x Q_1 - Y \cos(x) \, \partial_Y  Q_1
+ \left[ 1 - \cos(x) \right] Q_1 -  \partial^2_{YY} Q_{1} = - \sin(x) S_{11},
\label{qeq}
\eeq
a PDE whose inhomogeneity is a function of $S_{11}$ and whose solution has $Q_1 \sim \mathcal{O}(S_{11})$.  

\subsubsection{Solution of $S_{11}$}
In order to calculate $\tu_s$, we must compute $Q_1$, and therefore must explicitly compute $S_{11}$
from \eqref{s11}.  Here we show that equation \eqref{s11} has an exact  solution in the form
\beq
S_{11}(x,Y) = G(x) e^{-\frac{ F(x) Y^2}{2}} + H(x)
\label{s11form}
\eeq
so that
\beq
\begin{split}
\partial_Y S_{11} &= - Y G F  e^{-\frac{ F Y^2}{2}} \\
\partial_{YY} S_{11} & =  - G F \left[ 1 - Y^2 F  \right] e^{-\frac{ F Y^2}{2}} \\
\partial_x S_{11} & = \left[ G_x - \frac{Y^2 F_x G}{2} \right] e^{-\frac{ F Y^2}{2}}.
\end{split}
\eeq
 Substituting (\ref{s11form}) into (\ref{s11}) we find three ODEs for $F,G,H$.  The equation for $H$ decouples from the rest and absorbs the homogeneous term
 \beq
\sin(x) H_x + H \left[ 1 - 2 \cos(x) \right] - 1 = 0.
\label{h}
\eeq 
The equation for $F$ arises by setting the coefficient of  $Y^2 e^{-\frac{ F Y^2}{2}}$ to zero
\beq
F^2  - \cos(x) F + \frac{\sin(x) F_x}{2} = 0.
\label{f}
\eeq
This is a nonlinear,  inhomogeneous, first order equation for $F(x)$.   By setting the coefficient of
 $e^{-\frac{ F Y^2}{2}}$ to zero we arrive at a first order linear equation for $G(x)$
\beq
\sin(x) G_x + \left[ 1 - 2 \cos(x) \right] G + FG = 0.
\label{g}
\eeq
%

$H(x)$ will contribute a term to the solution of $Q_{1}$ but, since it is independent of $x$,  it will not contribute to the vorticity equation, \eqref{vort};  we do not record its solution.  
We can see by inspection that the solution to (\ref{f}) is
\beq
F(x) = \frac{ 1 + \cos(x)}{2} = \cos^2 \lp \frac{x}{2} \rp. 
\eeq

Note that the equation \eqref{g} is linear, meaning that any multiple of a solution remains a solution.  We can write \eqref{g} in the form
\beq
\frac{d (\ln(G))}{d x} = \frac{ 3( \cos(x) - 1)}{2 \sin(x)} = - \frac{3}{2} \cdot \frac{2 \sin^2\lp \frac{x}{2} \rp}{ 2 \sin \lp \frac{x}{2} \rp \cos \lp \frac{x}{2} \rp}
= -\frac{3}{2} \tan \lp \frac{x}{2} \rp
\eeq
with antiderivative
\beq
\ln(G(x)) = C +  3 \ln \lp \cos\lp \frac{x}{2} \rp \rp
\eeq
yielding
\beq
G(x) = S_0 \cos^3 \lp \frac{x}{2} \rp.  
\eeq
So we conclude that the asymptotic form of the primary component of the conformation tensor is
\beq
S_{11}(x,Y) = S_0 \cos^3 \lp \frac{x}{2} \rp 
e^{-\frac{1}{2} \cos^2 \lp \frac{x}{2} \rp Y^2 } + H(x).
\label{s11_sol}
\eeq
\subsubsection{Solution of $Q_1$}

Substituting $S_{11}$ from \eqref{s11_sol} into equation (\ref{qeq}) we now solve
\beq
\begin{split}
\sin(x) \, \partial_x Q_1 - Y \cos(x) \, \partial_Y  Q_1 &
+ \left[ 1 - \cos(x) \right] Q_1 -  \partial^2_{YY} Q_{1} = \\ &
 - \sin(x) \left[
 S_0 \cos^3 \lp \frac{x}{2} \rp 
e^{-\frac{1}{2} \cos^2 \lp \frac{x}{2} \rp Y^2 } + H(x)\right].
\label{qeq2}
\end{split}
\eeq
Again we  seek a solution of the form
\beq
Q_1 = M(x) e^{-\frac{1}{2} \cos^2 \lp \frac{x}{2} \rp Y^2 } + N(x).
\label{q1form}
\eeq
It is remarkable that the simple form (\ref{q1form}) provides an exact solution of (\ref{qeq2}), and we provide the details in order to convince the reader.  
After substituting \eqref{q1form} into \eqref{qeq2}, 
we find that the terms which are independent of $Y$ give an equation for $N(x)$
\beq 
\sin(x) \, \partial_x N + \left[ 1 - \cos(x) \right] N = -\sin(x) H(x)
\eeq
the solution of which requires  $H(x)$ from (\ref{h}).  
 Again, $N(x)$ is not needed since it does not affect the flow.   
 
The partial derivatives of the Gaussian terms are
\beq
\begin{split}
\partial_x Q_1 = & \left[ M_x + \frac{1}{2} Y^2  M \cos \lp \frac{x}{2} \rp 
\sin \lp \frac{x}{2} \rp \right] e^{-\frac{1}{2} \cos^2 \lp \frac{x}{2} \rp Y^2 }
\\
\partial_Y Q_1 = &
-  Y M  \cos^2 \lp \frac{x}{2} \rp e^{-\frac{1}{2} \cos^2 \lp \frac{x}{2} \rp Y^2 }
\\
\partial^2_{YY} Q_1 = & M \left[
Y^2 \cos^4 \lp \frac{x}{2} \rp  - \cos^2 \lp \frac{x}{2} \rp 
\right]
e^{-\frac{1}{2} \cos^2 \lp \frac{x}{2} \rp Y^2 }.
\end{split}
\label{q1_derivatives}
\eeq
Substituting the derivatives from \eqref{q1_derivatives} into equation (\ref{qeq2}) and collecting the
coefficients of the Gaussian, we find
\beq
\begin{split}
\sin(x) \left[ M_x + \frac{1}{2} Y^2  M \cos \lp \frac{x}{2} \rp 
\sin \lp \frac{x}{2} \rp \right] + \cos(x) Y^2   \cos^2 \lp \frac{x}{2} \rp M   & \\
+ \left [  1 - \cos \lp x \rp \right]  M
+ M  \cos^2 \lp \frac{x}{2} \rp  - Y^2 M  \cos^4 \lp \frac{x}{2}  \rp 
 = & \\  - \sin(x) S_0  \cos^3 \lp \frac{x}{2} \rp. &
\end{split}
\label{meq1}
\eeq
Simplifying with some trigonometric identities,
equation (\ref{meq1}) becomes
\beq
\begin{split}
2  \sin \lp \frac{x}{2} \rp  \cos \lp \frac{x}{2} \rp M_x  + 
 Y^2  M \left[ \cos \lp \frac{x}{2} \rp  \sin \lp \frac{x}{2} \rp \right]^2 + & \\
 Y^2  M \left[ \cos^4 \lp \frac{x}{2} \rp  -    \sin^2 \lp \frac{x}{2} \rp \cos^2  \lp \frac{x}{2} \rp \right] 
 +
 2    \sin^2 \lp \frac{x}{2} \rp  M +  & \\
   \cos^2 \lp \frac{x}{2} \rp  M - 
  Y^2 M  \cos^4 \lp \frac{x}{2}  \rp 
 = & \\  - 2 S_0  \sin \lp \frac{x}{2} \rp \cos^4 \lp \frac{x}{2} \rp .
\end{split}
\label{meq2}
\eeq
The terms multiplying $Y^2$ cancel one another, resulting  in a linear equation for $M(x)$
\beq
2  \sin \lp \frac{x}{2} \rp  \cos \lp \frac{x}{2} \rp M_x  +
\left[ 1 +   \sin^2 \lp \frac{x}{2} \rp  \right] M = 
- 2 S_0  \sin \lp \frac{x}{2} \rp \cos^4 \lp \frac{x}{2} \rp.
 \label{meq3}
 \eeq
whose solution is surprisingly simple,
 \beq
 M(x) = - S_0  \sin \lp \frac{x}{2} \rp \cos^2 \lp \frac{x}{2} \rp.
\eeq
Therefore, the dominant component of the stress divergence is 
\beq
Q_1 = - S_0  \sin \lp \frac{x}{2} \rp \cos^2 \lp \frac{x}{2} \rp 
e^{-\frac{1}{2} \cos^2 \lp \frac{x}{2} \rp Y^2 }+ N(x),
\label{q1final}
\eeq
where we emphasize that the stretched variable is defined to be
$Y= y / \sqrt{\nu} $.  
The solutions  \eqref{s11_sol} and \eqref{q1final} show that the stress and stress divergence are
localized within $\sqrt{\nu}$ of the $x$-axis.   In the far field, they are weighted Dirac $\delta$-distributions.   

\subsection{Solving for the stress induced velocity, $ \u_s$}
Substituting $Q_1$ from (\ref{q1final}) into the vorticity equation (\ref{om_s}) and using
$\Delta \psi_S = \omega_S$,
 the induced stream function solves the bi-laplacian equation
\beq
\Delta^2 \psi_S = \frac{1}{2} \partial_y Q_1 + O(\nu)
\label{stream}
\eeq
where
\beq
\Delta^2 = \left( \partial^2_{yy} + \partial^2_{xx} \right)^2.
\eeq
We will solve this equation in two different ways for the two different regions.
\begin{enumerate}
\item In the far field, we approximate the right hand side as the derivative of a $\delta$-distribution  and invert the bi-Laplacian on this source.
\item In the near field, we use an anisotropic scaling of the derivatives to simplify the operator on the left hand side.  The lowest order
terms retain only $Y$- derivatives. 
\end{enumerate}

\subsubsection{The outer approximation}
We define an intermediate variable
$ \psi_s = \partial_y \phi $
and solve the equation
\beq
\Delta^2 \phi = -\frac{S_0}{2} 
\sin \lp \frac{x}{2} \rp \cos^2 \lp \frac{x}{2} \rp 
e^{-\frac{ y^2 \cos^2 \lp x/2 \rp }{2\nu}  }
\label{phi_eq}
\eeq
using the original coordinate $y$ in the far field.  The $y$ length scale in the Gaussian is
\beq
L \sim \frac{ \sqrt{2 \nu} }{\cos \lp \frac{x}{2} \rp},
\eeq
which is $O(\sqrt{\nu})$ outside of $||x| - \pi|  < \sqrt{\nu}$.  Hence 
for almost all $x$, with $y > L,$  the right hand side of \eqref{phi_eq} is small.  Near $x = \pm \pi$, $L$ is no longer small and the 
 following approximations will fail.   However, for (almost all) points outside these turning points, we define
 the mollified Dirac $\delta$- distribution,  
\beq
\delta_L(y) = \frac{e^{-y^2/L}}{\sqrt{\pi\, L} }. 
\eeq
The smoothed $\delta$-distribution has the property that its integral is 1.   Multiplying and dividing by $\sqrt{\pi \, L}$ we write the $\phi$ equation as
\beq
\begin{split}
\Delta^2 \phi = & -S_0 \sqrt{ \frac{\pi \, \nu}{2} }  \, 
\sin \lp \frac{x}{2} \rp \cos \lp \frac{x}{2} \rp \; \delta_L(y) \\
= & -\frac{S_0}{2} \sqrt{ \frac{\pi \, \nu}{2} }  \, 
\sin \lp x \rp \; \delta_L(y),
\end{split}
\eeq
and solve for $\phi$ in the limit $L \rightarrow 0$.  Now we can seek a separable solution 
\beq \phi(x,y) = 
 -\frac{S_0}{2} \sqrt{ \frac{\pi \, \nu}{2} }  \, 
\sin \lp x \rp 
 \; \Phi(y),
\eeq
where 
\beq
\left( \frac{d^2}{dy^2} - 1 \right)^2 \Phi = \delta_0(y).
\label{anti_helm}
\eeq
The solution to equation \eqref{anti_helm} has $\Phi$ and its first two partial derivatives continuous at $y=0$.  The jump in the third derivative at $y=0$ is 
\beq
[ \Phi_{yyy}] = 1,
\eeq
which is solved by
\beq
\Phi = \frac{\left[ 1 + |y| \right] }{4} e^{- |y|},
\eeq
so that
\beq
\Phi_y = -\frac{y}{4} e^{- |y|},
\eeq
and
\beq
\psi_s = \frac{S_0}{8} \sqrt{ \frac{\pi \, \nu}{2} }  \, 
\sin \lp x \rp 
 \; y e^{- |y|}.
 \label{psis}
\eeq
Therefore, the induced velocity field  is
\beq
\left[ \begin{array}{c} u_s \\ v_s \end{array} \right] 
= -\frac{S_0}{8} \sqrt{ \frac{\pi \, \nu}{2} }  
\left[ \begin{array}{c} 
\sin(x) \, \lp 1 - |y| \rp 
\\ 
-y \cos(x)
\end{array} \right] e^{-|y|} . 
\label{us}
\eeq
%

The original hypothesis was that the induced velocity field, $\u_s$ should have the same functional form as the total velocity field, $\tu$ near where the stress is maximum.  
Clearly \eqref{us} and \eqref{u} do not have the same form everywhere.   However, the stress is sharply localized in a layer of width $L \sim \sqrt{\nu} \ll 1 $ near $y=0$, where the induced velocity is approximately
\beq
\left[ \begin{array}{c} u_s \\ v_s \end{array} \right] 
\approx -\frac{S_0}{8} \sqrt{ \frac{\pi \, \nu}{2} }  
\left[ \begin{array}{c} 
\sin(x)  \\ 
-y \cos(x)
\end{array} \right] + ...
\label{us_approx}
\eeq
which is proportional to $\tu$ from \eqref{u}.  

\subsubsection{The inner approximation}
\label{inner}
In order to get the correct solution of the velocity near and within the stress island, we must consider this
boundary layer region again using the scaled $y$ variable.   Upon substituting equation (\ref{q1final}) into (\ref{stream}), using the definition of $Y$,  retaining the largest terms on the left hand side, and integrating once with respect to $y,$ we find
\beq
\frac{1}{\nu^{3/2}} \partial^3_{YYY} \psi_s =  - \frac{S_0}{2}
  \sin \lp \frac{x}{2} \rp \cos^2 \lp \frac{x}{2} \rp  e^{-\frac{1}{2} \cos^2 \lp \frac{x}{2} \rp Y^2 }.
\eeq
Defining a scaled boundary layer variable
\beq
z = \frac{\cos \lp  \frac{x}{2} \rp }{\sqrt{2 }} Y,
\eeq
and the stream function equation becomes
\beq
\frac{d^3}{dz^3} \psi_s = -S_0 \sqrt{2 \nu^3} \tan  \lp \frac{x}{2} \rp  e^{-z^2}.
\eeq
After integrating three times, we arrive at the solution
\beq
\psi_s =   -S_0 \sqrt{2 \nu^3} \tan  \lp \frac{x}{2} \rp 
\left\{ \frac{\sqrt{\pi}}{8} \left[ 2 z^2 + 1\right] {\rm erf}(z) + \frac{z}{4} e^{-z^2} +  c_1 z   \right\},
\label{psis_inner}
\eeq
where $c_1$ is constant with respect to $Y$, although not necessarily with respect to $x$.  
In order to match the $y$-dependence of $\psi_s$, we evaluate $\psi_s$ from the outer solution in  
 equation \eqref{psis} in the limit $y\rightarrow 0,$ compare it to 
 $\psi_s$ in the inner solution, and substitute the definition 
 $ z = \frac{\cos \lp  \frac{x}{2} \rp  y}{\sqrt{2 \nu}}$.  
 
The large z limit of the near field solution is
\beq
\begin{split}
  \lim_{z \rightarrow \infty} \psi_s
& =  -S_0 \sqrt{2 \nu^3} \tan  \lp \frac{x}{2} \rp  
\left\{ \frac{\sqrt{\pi}}{4} z^2 + c_1 z \right\} \\
& = -S_0 \sqrt{2 \nu^3} \tan  \lp \frac{x}{2} \rp   
\left\{  \frac{\sqrt{\pi}}{4} \frac{\cos^2 \lp  \frac{x}{2} \rp }{2 \nu} \; y \, |y|
+ c_1 \frac{\cos \lp  \frac{x}{2} \rp }{\sqrt{2 \nu}} \; y
\right\} \\
& = -S_0 \sqrt{2 \nu^3} \sin  \lp \frac{x}{2} \rp   
\left\{  \frac{\sqrt{\pi}}{4} \frac{\cos \lp  \frac{x}{2} \rp }{2 \nu} \; y \, |y|
+ \frac{c_1 y }{\sqrt{2 \nu}} 
\right\}.
\end{split}
\eeq
The term which is quadratic in $y$ matches the outer solution automatically.  The linear term in $y$ matches the
outer solution if
\beq
c_1 = -\frac{\cos \lp \frac{x}{2} \rp }{4}\sqrt{\frac{\pi}{2 \nu}}.
\eeq

Near $y=0$, the near field approximation yields the horizontal velocity
\beq
\begin{split}
u_s &= - \partial_y \psi_s \\
& =  - \frac{\cos \lp  \frac{x}{2} \rp }{\sqrt{2 \nu}} \; \partial_z \psi_s \\
& \approx S_0 \nu \sin \lp \frac{x}{2} \rp \left\{ \frac{1}{4} + \frac{1}{4} + c_1 \right\} \\
& \approx S_0 \nu \sin \lp \frac{x}{2} \rp \left\{ \frac{1}{2}   -\frac{\cos \lp \frac{x}{2} \rp }{4}\sqrt{\frac{\pi}{2 \nu}} \right\} \\
& \approx -\frac{S_0}{8} \sqrt{\frac{\pi \nu}{2 } }\left[ \sin\lp x \rp  - 4 \sqrt{\frac{2 \nu}{\pi}} \sin\lp\frac{x}{2} \rp  \right].
\end{split}
\eeq
Notice a few features of the near field horizontal velocity.  First, we have only retained
terms independent of $y$ because we only need to study $u_s$ on the axis.  Next,  the first term matches the first term in the 
far field approximation and is the leading order term in the near field, while the second term is $\mathcal{O}(\sqrt{\nu})$ smaller.    The second term describes a shrinking of the horizontal 
extent of the horizontal velocity by  $\mathcal{O}(\sqrt{\nu}),$ i.e. the zonal velocity is zero on $y=0$ at $x_*$ where
\beq
\cos \lp \frac{x_*}{2} \rp = \sqrt{\frac{8 \nu}{\pi}} \Longrightarrow x_* \approx \pi - 4 \sqrt{\frac{2 \nu}{\pi}}.
\eeq
This further implies that a line of stress islands along $y=0$ will be drawn to one another by their tendency to pull their endpoints toward  their centers.

\subsubsection{Stress islands  on a periodic domain}
The stream function in (\ref{psis}) accounts for stress islands on an $x$-periodic, but not $y$-periodic, domain.   In order to account for $y-$periodicity, we need to solve (\ref{anti_helm}) with a periodized $\delta$ function.  Solutions to the periodic version of  (\ref{anti_helm}) on $ 0 < y < 2 \pi$ are exponentials and $y$ times exponentials, which can be written
\beq
\Phi = A \cosh(y-\pi) +B (y-\pi) \sinh(y-\pi)  
+ C \sinh(y - \pi) + 
D (y-\pi) \cosh(y-\pi).
\eeq
The function is centered at $y=\pi$ in order to exploit the symmetry around this point.  In fact, $\Phi$ should be symmetric around this point (two derivatives of $\Phi$ are proportional to $u$) so we can simplify the expression
\beq
\Phi = A \cosh(y-\pi) + B (y-\pi) \sinh(y-\pi) .
\label{phipform}
\eeq
Establishing periodicity in the $y$-direction requires
\beq
\begin{split}
\Phi(2\pi) = \Phi(0), \quad & \quad 
\Phi_{yy}(2\pi) = \Phi_{yy}(0), \\
\Phi_{y}(2\pi) = \Phi_{y}(0), \quad & \quad 
\Phi_{yyy}(2\pi) = \Phi_{yyy}(0) - 1.
\end{split}
\eeq
The choice of symmetric $\Phi$
means that its second derivative automatically satisfies
the periodicity conditions, and one need only use the first and third derivative
conditions.     
The first derivative is
\beq
\Phi_y 
 = (A + B)\sinh(y-\pi) + B (y-\pi) \cosh(y-\pi),
\eeq
which, when enforcing periodicity results in
\beq
A = -B (1 + \pi \coth(\pi)),
\eeq
so that
\beq
\begin{split}
\Phi &=
 B \left[ (y-\pi) \sinh(y-\pi) - \lp 1 + \pi \coth(\pi) \rp \cosh(y - \pi ) \right] \\
\Phi_y & = 
B \left[ (y-\pi) \cosh(y-\pi) - \pi \coth(\pi)  \sinh(y - \pi ) \right] 
 \\
\Phi_{yy} & = B \left[\lp 1 - \pi \coth(\pi) \rp \cosh(y-\pi) + (y-\pi) \sinh(y-\pi)\right]
 \\
\Phi_{yyy} & =B \left[\lp 2 - \pi \coth(\pi) \rp \sinh(y-\pi) + (y-\pi) \cosh(y-\pi)\right].
\label{phipform2}
\end{split}
\eeq
Lastly, using the third derivative condition yields
\beq
B = \frac{-1}{4 \, \sinh(\pi)}.
\eeq
Using $\Phi_y$ to construct $\psi_{P}$ (P  is used to denote ``periodic'') we find
\beq
\psi_P = \frac{S_0 \, \coth(\pi) }{8 } \sqrt{\frac{\pi \nu}{2}} \, \sin(x) \,
\left[ (y-\pi)\, \frac{\cosh(y-\pi)}{\cosh(\pi)}  - \pi \,\frac{\sinh(y - \pi )}{\sinh(\pi)} \right],
\label{psip}
\eeq
on $0 < y < 2 \pi$. It must be periodically extended outside of this domain, for example to $-2\pi < y < 0$,
\beq
\psi_P = 
 \frac{S_0 \, \coth(\pi) }{8 } \sqrt{\frac{\pi \nu}{2}} \, \sin(x) \,
\left[ (y + \pi)\, \frac{\cosh(y + \pi)}{\cosh(\pi)}  - \pi \,\frac{\sinh(y + \pi )}{\sinh(\pi)} \right].
\label{psipneg}
\eeq

\subsubsection{Multiple stress islands on a doubly periodic domain}
\label{per}

\begin{figure}
        \centering
                \includegraphics[width=1.05\textwidth]{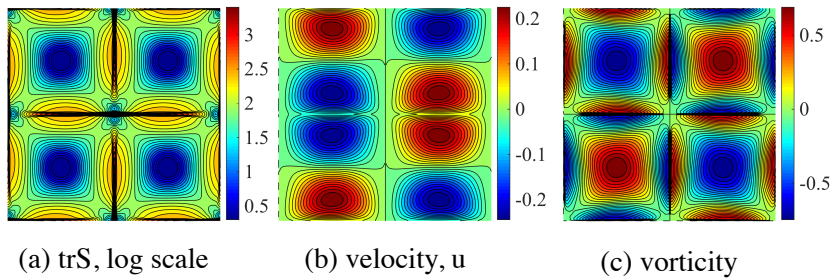}
           \vspace{-2.2in}             \caption{$4-$roll mill simulations on $[0,2\pi]^2$ with $\nu=0.00025,$ $\Wi = 20$}\label{fig:4roll}
\end{figure}

In \cite{TS2007,T2011}, a doubly periodic  $4$-roll mill type geometry was studied on $[0,2\pi]\times[0,2\pi].$  The background force $\f=(2\sin x\cos y,-2\cos x\sin y)$ prescribed a flow with an extensional point at the origin,  stretching in the $x$-direction and squeezing in the $y$-direction.
  The results of numerical simulations of this flow for $\Wi=20,$ $\nu=0.00025$
  are plotted in  in Fig. \ref{fig:4roll} after a time $t=5\Wi$ when the
  flow has equilibrated.    Fig.\ref{fig:4roll} (a) shows contours of the trace of the conformation tensor $S_{11}+S_{22},$ on a log-scale, (b) shows the first component of the velocity $u,$ and (c) the vorticity $\d_xv-\d_yu.$ Unlike our theoretical background flow, here there are four stress islands contributing to the flow in the domain $0<x<\pi,0<y<\pi.$  In order to compare our theoretical predictions with this flow geometry we need to account for all of these stress islands, we will show results of the comparison in Sec. \ref{numcomp}  

For the island oriented parallel to the x-axis located at $(x_1,y_1) = (\pi, \pi)$, we need to use the expression in (\ref{psipneg}), but shift its location,
\beq
\begin{split}
\psi_1 = & \, 
 \frac{S_0 \, \coth(\pi) }{8 } \sqrt{\frac{\pi \nu}{2}} \, \sin(x-x_1) \, \\
 &
\left[ (y-y_1 + \pi) \, \frac{\cosh(y-y_1 + \pi)}{\cosh(\pi)}  - \pi \,\frac{\sinh(y-y_1 + \pi)}{\sinh(\pi)} \right] \\
= & \, 
-  \frac{S_0 \, \coth(\pi) }{8 } \sqrt{\frac{\pi \nu}{2}} \, \sin(x) \,
\left[ y\, \frac{\cosh(y)}{\cosh(\pi)}  - \pi \,\frac{\sinh(y)}{\sinh(\pi)} \right] .
\label{psip1}
\end{split}
\eeq

The other two islands are rotated by $\pi/2$, which is effected by the replacement $(x,y) \rightarrow (y,-x)$.  The island  along the y-axis is located at $(x_2,y_2) = (0,\pi)$ and from its perspective, the first quadrant is ``below'' it, so we again use the expression in (\ref{psipneg}).  Therefore
\beq
\begin{split}
\psi_2 = & \,\frac{S_0 \, \coth(\pi) }{8 } \sqrt{\frac{\pi \nu}{2}} \, \sin(y-y_2) \, \\
 &
\left[ (-(x-x_2) + \pi)\, \frac{\cosh(-(x-x_2) + \pi)}{\cosh(\pi)}  - \pi \,\frac{\sinh(-(x-x_2) + \pi )}{\sinh(\pi)} \right]
\\
= & \, 
 \,\frac{S_0 \, \coth(\pi) }{8 } \sqrt{\frac{\pi \nu}{2}} \, \sin(y) \, \left[ (x - \pi)\, \frac{\cosh(x - \pi)}{\cosh(\pi)}  - \pi \,\frac{\sinh(x - \pi )}{\sinh(\pi)} \right].
\label{psip2}
\end{split}
\eeq
Finally we consider the island at $(x_3,y_3) = (\pi,0)$; from the perspective of this stresslet the first quadrant is ``above'' it, meaning that the expression in (\ref{psip}) is relevant,
\beq
\begin{split}
\psi_3 = & \,\frac{S_0 \, \coth(\pi) }{8 } \sqrt{\frac{\pi \nu}{2}} \, \sin(y-y_3) \, \\
 &
\left[ (-(x-x_3) - \pi)\, \frac{\cosh(-(x-x_3) - \pi)}{\cosh(\pi)}  - \pi \,\frac{\sinh(-(x-x_3) - \pi )}{\sinh(\pi)} \right]
\\
= & \, -
\frac{S_0 \, \coth(\pi) }{8 } \sqrt{\frac{\pi \nu}{2}} \, \sin(y) \, 
\left[ x\, \frac{\cosh(x)}{\cosh(\pi)}  - \pi \,\frac{\sinh(x)}{\sinh(\pi)} \right].
\label{psip3}
\end{split}
\eeq

Notice that each of the $\psi$ expressions is positive in $(x,y) \in [0,\pi]\times [0.\pi]$,  which means that they are additive, i.e., the induced flow from each of the  four stress islands reinforce the flow from the others.  Notice also that $\psi_1$ and $\psi_2$ are the same expression with $x$ and $y$ interchanged, as are $\psi_1$ and $\psi_3$.   

The sum of the four expressions  yields
\beq
\psi_{tot} =  \; \frac{S_0 \, \coth(\pi) }{8 } \sqrt{\frac{\pi \nu}{2}} \, \\
\left[\sin(x) M(y)+\sin(y) M(x)\right],\label{psitotper}
\eeq
where the function
 \beq M(x)\equiv\left[
\frac{(x-\pi)\cosh(x-\pi) - x \cosh(x) }{\cosh(\pi)} - 
\pi \frac{\sinh(x - \pi )- \sinh(x)}{\sinh(\pi)}
\right].\label{mess}\eeq

\subsection{Calculating $S_0$}
Near $y=0$, we expect the {\bf total} velocity,
\beq
\left[ \begin{array}{c} u \\ v \end{array} \right]
= 
\left[ \begin{array}{c} \sin(x)  \\ -y \cos(x) \end{array} \right]
\eeq
to  consist of an externally  forced portion
\beq
\left[ \begin{array}{c} u_f \\ v_f \end{array} \right] 
= \Wi \, 
\left[ \begin{array}{c} \sin(x)  \\ -y \cos(x) \end{array} \right],  
\eeq
and a stress induced portion
\beq
\left[ \begin{array}{c} u_S \\ v_S \end{array} \right] 
= \lp 1 - \Wi \rp  \, 
\left[ \begin{array}{c} \sin(x)  \\ -y \cos(x) \end{array} \right]. 
\label{us1} 
\eeq
We need to compare this form of the velocity field with the solution in equation (\ref{psitotper}) evaluated near the origin.  
The function $M(x)$ in equation \eqref{mess} can be approximated as $1.144\sin(x),$ near $x=0,$ so near the origin
\beq
\psi_{tot} \approx 
 \; \frac{S_0 \, \alpha }{8 }  \sqrt{\frac{\pi \nu}{2}} \sin(x) \sin(y),
 \label{psitot_last}
 \eeq
where $\alpha = 2 \coth(\pi) M'(0)  \approx 2 \, \coth(\pi) \times 1.144 \approx 2.297$. 
Comparing \eqref{psitot_last} with \eqref{us1} we find
\beq
\frac{S_0}{8} \sqrt{ \frac{\pi \, \nu}{2} }  \alpha =  \Wi - 1
\quad
{\rm or} \quad
S_0 = \frac{8 \lp \Wi - 1 \rp}{\alpha}   \sqrt{ \frac{2}{\pi \, \nu} } .\label{S0theo}
\eeq
In Section \ref{numcomp} we will compare this theoretical prediction to simulations of the Stokes-Oldroyd-B system.   In the case of the singly periodic velocity field $\alpha = 1$, however we do not have any numerical simulations with which to compare this result.  

\subsection{The correctly scaled velocity field.}
The original rescaling of the velocity field by the Weissenberg number
was simply a computational convenience.  We must remove this scaling in order to get the actual velocity field
\beq
\u = \frac{\tu}{\Wi}.
\eeq
In the case of the singly periodic velocity field, the background velocity,
\beq
\left[ \begin{array}{c} u_f \\ v_f \end{array} \right] 
=  
\left[ \begin{array}{c} \sin(x)  \\ -y \cos(x) \end{array} \right]\label{uf2}
\eeq
is independent of the Weissenberg number, as it must be from equation \eqref{momcon}.  The stress induced velocity field is
\beq
\left[ \begin{array}{c} u_s \\ v_s \end{array} \right] 
= \lp \frac{1}{\Wi} - 1 \rp 
\left[ \begin{array}{c} 
\sin(x) \, \lp 1 - |y| \rp 
\\ 
-y \cos(x)
\end{array} \right] e^{-|y|},
\label{us2}
\eeq
so that  the total stream function is
\beq
\psi = - y \sin(x) \left[ 1 + \lp \frac{1}{\Wi} - 1 \rp e^{-|y|} \right].
\eeq
Near the stress island, the velocity scales as the reciprocal of $\Wi,$ but far away the stress induced response decays.  The same rescaling yields the stream function for the doubly periodic example,
\beq
\psi = -\sin(x) \sin(y) + \lp 1 - \frac{1}{\Wi} \rp  
\left[\frac{\sin(x) M(y) + \sin(y) M(x) }{ 2 M'(0)}\right].
\eeq

\section{Comparison to Numerical Simulations}\label{numcomp}

We compare our theoretical results to numerical simulations of the Stokes-Oldroyd-B system defined in Eqs.\eqref{momcon}-\eqref{OB} where $\f$ creates a doubly periodic $4-$roll mill type geometry,  \beq \f = (2\sin x\cos y,-2\cos x\sin y).\eeq  The system is solved in a doubly periodic domain using a pseudo-spectral method (\cite{TS2007,T2011}).   These simulations were performed with $\triangle x=2\pi/N,$ for $N=2^{9},$ with $\nu=0.0125,$ $0.0025,$ $0.00125,$ $0.00025,$ where $\nu/(\triangle x)^2\approx 83,$ $17,$ $8,$ $2.$


\begin{table}[h!]\begin{center}
\begin{tabular}{|c||c|c|c|}
\hline 
$\nu$ & $\Wi = 10$ & $\Wi = 15$ & $\Wi =20$\\\hline\hline
0.01250 & 0.08250 & 0.04338 & 0.00613  \\\hline
0.00250 & 0.02542 & 0.01073 & 0.0035 \\\hline
0.00125 & 0.00785 & 0.01107 & 0.00575\\\hline
0.00025 & 0.04153 & 0.00694 & 0.00322 \\\hline
\end{tabular}\caption{Relative difference in theoretical maximum, $S_{11}(0,0)$, compared with $4-$roll mill simulation. }\label{Table:reldiff}
\end{center}\end{table}

In Table \eqref{Table:reldiff} we show the relative difference between the theoretical prediction in Eq. \eqref{S0theo} and the maximum of the first component of the stress tensor at equilibration for the $4-$roll mill simulations, using  the value $\alpha = 2.297\approx 2\coth(\pi)\times 1.144,$ in our calculation of $S_0$ from equation (\ref{S0theo}).   
 The error in the approximation is smallest for $\Wi = 10,20 $ for $\nu=0.00125,$ and for $\Wi = 15$ is smallest for $\nu=0.00025,$ where the error $\lesssim 1\%$.  We see that we can get excellent matching between the theoretical prediction and the numerical solutions when the diffusion is approximately $2-8$ times $(\triangle x)^2.$

Decreasing the diffusion without increasing resolution will increase the error in the numerical simulation since we need a grid size
$\triangle x < \sqrt{\nu}/C$, for some number $C > 2$, which is independent of $\nu$.  
On the other hand, increasing $\nu$  increases the error in the asymptotic approximation.  Therefore the asymptotic approximation becomes more valid at small values of the stress diffusion - which is precisely where high resolution is needed in numerical methods, thereby greatly increasing computational time.

%

\begin{figure}[h]
  \centering\vspace{-1in}
    \includegraphics[width=0.8\textwidth]{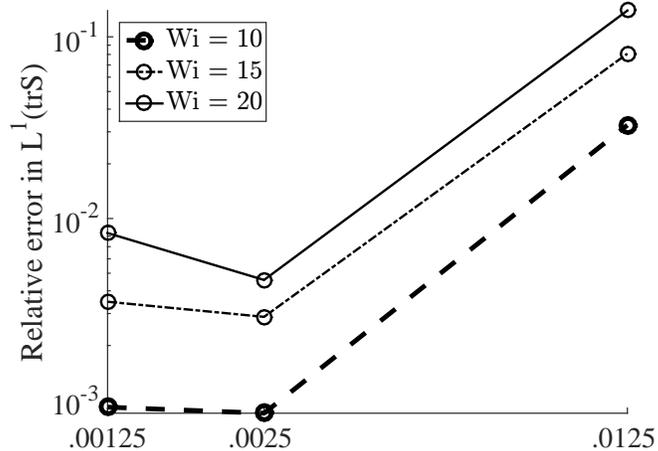}\vspace{-.35in}\label{fig:L1trS}  \vspace{-1in}\caption{Relative error in $L^1(\textrm{tr}\S)$ as a function of $\nu,$ comparing with solution at $\nu=0.00025.$}\label{fig:L1trS}
\end{figure}

In our asymptotic solution, the maximum of the conformation tensor scales with $\frac{1}{\sqrt{\nu}}.$ 
While we do not expect $S_{11}(0,0)$ to converge as $\nu\rightarrow 0$ we do, however, expect the conformation tensor to converge in an integral-norm;  it is precisely this integral quantity that is necessary to find the solution for the velocity. Simulations in the $4-$roll geometry show that the integral is converging. 
In Fig. \ref{fig:L1trS} we plot the relative error in the $L^1-$norm of the trace of the conformation tensor, defined by  \[L^1(\textrm{tr}\S)\equiv\int_{0}^{2\pi}\int_0^{2\pi}|\textrm{tr}\S|\;dx \;dy,\] where we use the solution with diffusion coefficient $\nu=0.00025$ as the ``true" solution.  The maximum of the conformation tensor may be growing as $\nu\rightarrow 0,$ but the integral of this quantity converges as $\nu\rightarrow 0,$ and in fact for the values considered, the relative size is changing by less than one percent.  

\begin{figure}
        \centering\vspace{-.75in}
              \includegraphics[width=.5\textwidth]{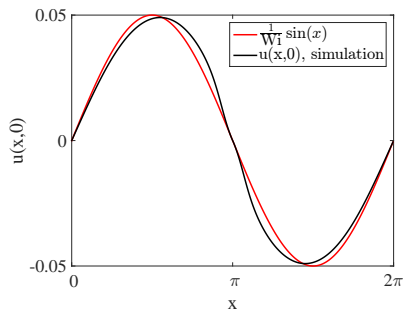}\vspace{-.75in}
          \caption{Comparison of simulation $u(x,0)$ with theoretically predicted $\frac{1}{\Wi}\sin(x)$, simulations use $\Wi=20,$ $\nu = 0.00025$}
          \label{fig:4rollslice_vel}
\end{figure}

It is not just at the extensional stagnation point where we see good agreement between theory and simulation. In a strip around the $y-$axis (and by symmetry along all the directions of stretching and compression in the 4-roll mill) the asymptotic theory captures the lowest order behavior of the velocity. For example we compare the theoretically predicted horizontal velocity along $y=0$ with the simulation.  
From equations  \ref{uf2}--\ref{us2}, the former is given by 
 \[u(x,y)|_{y=0}=\sin (x) +\lp \frac{1}{\Wi} - 1 \rp
\sin(x) \, \lp 1 - |y| \rp e^{-|y|}=\frac{1}{\Wi}\sin(x).\]
\begin{table}[h!]\begin{center}
\begin{tabular}{|c||c|c|c|}
\hline 
$\nu$ & $\Wi = 10$ & $\Wi = 15$ & $\Wi =20$\\\hline\hline
0.01250 & 0.03878 & 0.03397 & 0.02847  \\\hline
0.00250 & 0.04802 & 0.04374 & 0.03299 \\\hline
0.00125 & 0.04814 & 0.04483 & 0.03823 \\\hline
0.00025 & 0.04577 & 0.03675 & 0.02927  \\\hline
\end{tabular}\caption{Relative difference between theoretical and simulation in $u(\pi/2,0)$ compared with $4-$roll mill simulation. }\label{Table:reldiffvel}
\end{center}\end{table} 
Figure \ref{fig:4rollslice_vel} shows plots of both $\frac{1}{\Wi}\sin(x)$ and simulation results of $u(x,0)$ for $\Wi = 20,$ $\nu=0.00025.$  Results of relative error in the approximation at $x=\pi/2$ for a range of $\Wi,$ and $\nu$ are given in Table \ref{Table:reldiffvel}.  This approximation
 is valid for all $\nu$ in the asymptotically small limit since  this solution does not depend on diffusion. The error remains small in a strip around the axes of compression and extension near the stagnation point for $|y|\lesssim 0.1.$

\begin{figure}\vspace{-1in}
        \centering
               \hspace{-.5in} \includegraphics[width=1.05\textwidth]{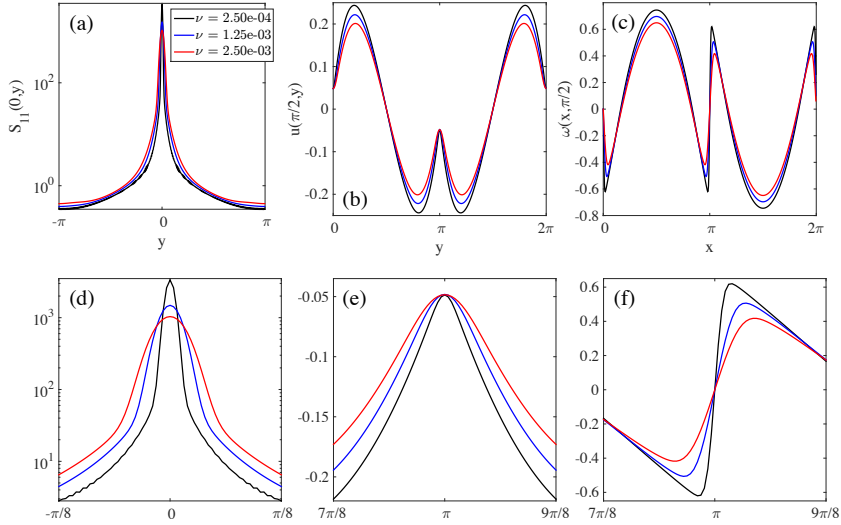}\vspace{-.75in}
          \caption{$4-$roll mill simulations for $\Wi = 20,$ for $\nu = 0.0025, 0.00125, 0.00025$ (a) principal component of conformation tensor along axis of compression $S_{11}(0,y)$ (b) velocity $u(\pi/2,y)$ (c) vorticity $\omega(x,\pi/2)$ (d)-(f) close-ups of (a)-(c) respectively}\label{fig:4rollslices}
\end{figure}

Now let us examine the dependence of the numerical solution on the stress diffusion, $\nu,$ in the $4-$roll mill numerical simulations.    Figure \ref{fig:4rollslices} (a) shows the first component of the conformation tensor along the $y$-axis, in the direction of compression and  Fig. \ref{fig:4rollslices} (d) is a close-up of (a). The Gaussian structure is evident, and looking at the close-up of the conformation tensor (d) we see a sharp second derivative near the origin located where 
the solution transitions from the singular behavior of the Gaussian, to the smoother behavior at the center of the $4$-roll mill (where the flow extension is weak).

 Figure \ref{fig:4rollslices} (b) and close-up (e) show the horizontal velocity in a vertical cut through the center of the $4$-roll mill, $x = \pi/2$.  Near $y=\pi$ the stress tensor is again singular.  The dip in the horizontal velocity is distinctly evident in both figures.  As described above, this dip is due to the response of the horizontal velocity to the stress island.   The strength of the velocity at the center of the stress island, $(x,y) = (\pi/2,\pi)$ is not sensitive to $\nu$ and
 clearly attains the value $u = \Wi^{-1} = .05$ at the center of the dip, as is predicted in the asymptotic theory.

  Finally,  Fig. \ref{fig:4rollslices} (c) and close-up (f), show that the vorticity has a jump which is smoothed out by diffusion, but the magnitude of the jump  converges with decreasing diffusion and in the limit of $\nu=0$, the asymptotic model predicts  \beq
\lim_{x \rightarrow \pi-} \omega(x,\pi/2) = 
\left(  1 -  Wi^{-1} \right) \frac{M''(0)}{2 M'(0)} \approx -0.775.  
\eeq
  Extrapolating the numerically computed $\omega$ linearly to $x = \pi^-$ yields $\omega \approx -0.725$, in excellent agreement with the asymptotic prediction.

  Matching the stress island to the center of the roll is a more difficult problem, and since the $4-$roll mill is a toy  geometry, performing the detailed calculations necessary to do the matching is unlikely to yield further insight.  
However, near the center of the roll, the flow is purely rotational and is dominated by the antisymmetric part of the velocity gradient matrix.
Therefore,  irrespective of the Weissenberg number, the flow near the center of the $4$-roll mill behaves like the low $\Wi$ limit because the symmetric part of the deformation tensor, which stretches the flow and creates the stress islands, is small.

 We posit that the stress tensor can be written $\S=\S_R+\S_S$
 where $\S_R$ denotes the regular solution, which dominates in the middle of the rolls, and $\S_S$ denotes the singular solution which describes the stress islands (and which we have already discussed in detail).   
Using the intuition that the flow at the center of the roll behaves like the
 low $\Wi$ flow locally, then $\S_R\approx\Id+\Wi(\na\u+\na\u^T)$
  \cite{TS2007}.
 Taking the curl of Eq. \eqref{momcon}, and substituting  $\S=\S_R+\S_S$ gives  \beq
  \Delta \omega = - \frac{\Delta \omega}{2} - 4 \sin(x) \sin(y).
  \label{torque1}
  \eeq
This gives a value for $\lap\omega$ at the center of the 4-roll mill,
\beq
- \Delta \omega = \frac{8}{3} \quad \Longrightarrow
\quad
- \frac{\Delta \omega}{2} = \frac{4}{3}.
\eeq For all $\nu=0.0125,$ $0.025,$ $0.00125,$ $0.00025$ and $\Wi = 10,$ $15,$ $20$ the error in this value at the center of the roll is less than $0.04\%$.  Although this solution is obtained in the low $\Wi$ limit, it is valid for all $\Wi.$  Note that near the axes of extension and compression this regular solution only has terms that are lower order in our expansion and can be ignored. 

Finally, we make a comment about using diffusion to enforce finite-extension. In the UCM model of viscoelastic fluids, an asymptotic scaling argument was used to show that at extensional points the width of a birefringent strand in the FENE-P model scales as $\frac{1}{\ell}+\frac{1-\ln \ell}{2\ell}\frac{1}{\Wi}+\mathcal{O}(\frac{1}{\Wi^2}) $ \cite{becherer2008scaling}.  As the Gaussian full width at half-maximum given in $S_0$ scales like $\sqrt{\nu}$ which in turn scales like $\frac{1}{\ell^2},$  we see that the birefringent strand constructed using diffusion will be much thinner than a corresponding strand using FENE-P.  The FENE-P model still requires some diffusion to evolve to steady state, in order to resolve the corners that arise in the cut-off of $\tr\S$ that arises at extensional stagnation points in FENE-P \cite{TS2007}.  It may be possible to use far less diffusion to regularize FENE-P and obtain accurate solutions.  An argument like we made above would be more complicated in that case, but if possible, it would be an important result.

\section{Conclusions}

We have found an analytic, asymptotic approximation of the steady-state equations for viscoelastic creeping flow in a neighborhood of an extensional stagnation point.  This approximation uses polymer stress diffusion as a regularization to find a solution in the form of a Gaussian for the principle component of the polymer stress in the stretching direction. 

At the extensional point, the stress becomes localized and highly stretched in a region of the outgoing streamlines of the stagnation point.  The Gaussian structure of the solution was recognized in \cite{T2011}, but without including the $x-$dependence in the solution it is not possible to get any information about the feedback to the velocity.  This paper used the special structure of the equations that arises when $u_y=0,$ which allows solutions to be obtained in orders of the asymptotic parameter, $\nu.$  The singular solutions capture the behavior of the elastic stress near the stagnation point, and can be used to find an approximation for the velocity response near the stagnation point. 

Due to the special structure of the equations this solution for the velocity is {\em independent} of the diffusion parameter.  This shows that the exact nature of the elastic stress at the extensional point is {\em not essential} to 
determine the behavior of the flow near the stagnation point. This is an important observation since many of the modifications to Oldroyd-B that are designed to incorporate finite extension or other rheological properties, such as FENE-P, Giesekus, PTT, inherit the difficulties of Oldroyd-B near extensional points.  This indicates that a small amount of diffusion, chosen carefully 
to depend on both the parameters of the flow and on how the flow turns around, can be used to determine an appropriate grid-size and diffusion for the problem that will exhibit sufficient smoothness as well as the ability to stretch to a physically valid length. 

This solution also gives an essential theoretical piece of the physical explanation for the instabilities in viscoelastic fluids that may lead to a deeper understanding of elastic turbulence.  We have shown that the 
velocity response to large stress is to decrease the vorticity near the regions of large stress which in turn leaves room for stress to grow.  Thus stress expels vorticity, which in turn creates stress.


%
%

%


\vspace{.1in}

\noindent J.A.B. was partially supported by NSF DMS 1009959 and 1313477.

\bibliographystyle{elsarticle-num}
\bibliography{BT_jnnfm}

\begin{thebibliography}{10}
\expandafter\ifx\csname url\endcsname\relax
  \def\url#1{\texttt{#1}}\fi
\expandafter\ifx\csname urlprefix\endcsname\relax\def\urlprefix{URL }\fi
\expandafter\ifx\csname href\endcsname\relax
  \def\href#1#2{#2} \def\path#1{#1}\fi

\bibitem{ATDG2006}
P.~E. Arratia, C.~Thomas, J.~Diorio, J.~Gollub, Elastic instabilities of
  polymer solutions in cross-channel flow, Physical review letters 96~(14)
  (2006) 144502.

\bibitem{soulages2009investigating}
J.~Soulages, M.~Oliveira, P.~Sousa, M.~Alves, G.~McKinley, Investigating the
  stability of viscoelastic stagnation flows in t-shaped microchannels, Journal
  of Non-Newtonian Fluid Mechanics 163~(1) (2009) 9--24.

\bibitem{liu2012oscillations}
B.~Liu, M.~Shelley, J.~Zhang, Oscillations of a layer of viscoelastic fluid
  under steady forcing, Journal of Non-Newtonian Fluid Mechanics 175 (2012)
  38--43.

\bibitem{haward2013instabilities}
S.~Haward, G.~McKinley, Instabilities in stagnation point flows of polymer
  solutions, Physics of Fluids (1994-present) 25~(8) (2013) 083104.

\bibitem{sousa2015purely}
P.~Sousa, F.~Pinho, M.~Oliveira, M.~Alves, Purely elastic flow instabilities in
  microscale cross-slot devices, Soft matter 11~(45) (2015) 8856--8862.

\bibitem{harris1993start}
O.~Harris, J.~Rallison, Start-up of a strongly extensional flow of a dilute
  polymer solution, Journal of non-newtonian fluid mechanics 50~(1) (1993)
  89--124.

\bibitem{harris1994instabilities}
O.~Harris, J.~Rallison, Instabilities of a stagnation point flow of a dilute
  polymer solution, Journal of non-newtonian fluid mechanics 55~(1) (1994)
  59--90.

\bibitem{poole2007purely}
R.~Poole, M.~Alves, P.~Oliveira, Purely elastic flow asymmetries, Physical
  review letters 99~(16) (2007) 164503.

\bibitem{thomases2009transition}
B.~Thomases, M.~Shelley, Transition to mixing and oscillations in a stokesian
  viscoelastic flow, Physical review letters 103~(9) (2009) 094501.

\bibitem{xi2009mechanism}
L.~Xi, M.~D. Graham, A mechanism for oscillatory instability in viscoelastic
  cross-slot flow, Journal of Fluid Mechanics 622 (2009) 145--165.

\bibitem{thomases2011stokesian}
B.~Thomases, M.~Shelley, J.-L. Thiffeault, A stokesian viscoelastic flow:
  Transition to oscillations and mixing, Physica D: Nonlinear Phenomena
  240~(20) (2011) 1602--1614.

\bibitem{giesekus1966}
H.~Giesekus, Die elastizit{\"a}t von fl{\"u}ssigkeiten, Rheologica Acta 5~(1)
  (1966) 29--35.

\bibitem{thien1977new}
N.~P. Thien, R.~I. Tanner, A new constitutive equation derived from network
  theory, Journal of Non-Newtonian Fluid Mechanics 2~(4) (1977) 353--365.

\bibitem{peterlin1961}
A.~Peterlin, Streaming birefringence of soft linear macromolecules with finite
  chain length, Polymer 2 (1961) 257--264.

\bibitem{BHAC1980}
R.~B. Bird, O.~Hassager, R.~Armstrong, C.~Curtiss, Dynamics of Polymeric
  Liquids, Vol. 2: Kinetic Theory, John Wiley and Sons, 1980.

\bibitem{OP2002}
R.~G. Owens, T.~N. Phillips, Computational rheology, Vol.~2, World Scientific,
  2002.

\bibitem{guy2015computational}
R.~D. Guy, B.~Thomases, Computational challenges for simulating strongly
  elastic flows in biology, in: Complex Fluids in Biological Systems, Springer,
  2014, pp. 361--400.

\bibitem{T2011}
B.~Thomases, An analysis of the effect of stress diffusion on the dynamics of
  creeping viscoelastic flow, J. Non-Newt. Fluid Mech 166 (2011) 1221--1228.

\bibitem{constantin2012note}
P.~Constantin, M.~Kliegl, Note on global regularity for two-dimensional
  oldroyd-b fluids with diffusive stress, Archive for Rational Mechanics and
  Analysis (2012) 1--16.

\bibitem{Larson1999}
R.~G. Larson, The structure and rheology of complex fluids, Vol.~2, Oxford
  university press New York, 1999.

\bibitem{KL1989}
A.~W. El-Kareh, L.~G. Leal, Existence of solutions for all {D}eborah numbers
  for a non-{N}ewtonian model modified to include diffusion, J. Non-Newton.
  Fluid Mech. 33 (1989) 257.

\bibitem{SB1995}
R.~Sureshkumar, A.~N. Beris, Effect of artificial stress diffusivity on the
  stability of numerical calculations and the flow dynamics of time-dependent
  viscoelastic flows, Journal of Non-Newtonian Fluid Mechanics 60 (1995) 53 --
  80.

\bibitem{lyazid1980velocity}
A.~Lyazid, O.~Scrivener, R.~Teitgen, Velocity field in an elongational polymer
  solution flow, in: Rheology, Springer, 1980, pp. 141--148.

\bibitem{gardner1982photon}
K.~Gardner, E.~Pike, M.~Miles, A.~Keller, K.~Tanaka, Photon-correlation
  velocimetry of polystyrene solutions in extensional flow fields, Polymer
  23~(10) (1982) 1435--1442.

\bibitem{rabin1986flow}
Y.~Rabin, F.~S. Henyey, D.~B. Creamer, Flow modification by polymers in strong
  elongational flows, The Journal of chemical physics 85~(8) (1986) 4696--4701.

\bibitem{harlen1990high}
O.~Harlen, J.~Rallison, M.~Chilcott, High-deborah-number flows of dilute
  polymer solutions, Journal of Non-Newtonian Fluid Mechanics 34~(3) (1990)
  319--349.

\bibitem{TS2007}
B.~Thomases, M.~Shelley, Emergence of singular structures in {O}ldroyd-{B}
  fluids, Phys. Fluids 19 (2007) 103103.

\bibitem{becherer2008scaling}
P.~Becherer, A.~N. Morozov, W.~v. Saarloos, Scaling of singular structures in
  extensional flow of dilute polymer solutions, Journal of Non-Newtonian Fluid
  Mechanics 153~(2) (2008) 183--190.

\end{thebibliography}

\end{document}